\documentclass[aps,prl,twocolumn,showpacs,superscriptaddress]{revtex4-1}  
\usepackage{graphicx}  
\usepackage{orcidlink}
\usepackage{dcolumn}   
\usepackage{bm}        
\usepackage{amssymb}   
\usepackage{lipsum}
\usepackage{xcolor}
\usepackage{amsmath}
\usepackage{overpic}
\usepackage{hyperref}
\usepackage{lineno}
\usepackage{slashed}
\usetikzlibrary{calc, positioning}
\usepackage[utf8]{inputenc}

\begin{document}
\begin{tikzpicture}[remember picture, overlay]
\node[anchor=north east] at ($(current page.north east)-(2,1)$){Belle II Preprint 2024-005};\\                    
\node[anchor=north east] at ($(current page.north east)-(2,1.5)$){KEK Preprint 2023-52 };
\end{tikzpicture}
\title{Search for Rare $b \to d\ell^+\ell^-$ Transitions at Belle}
  \author{I.~Adachi\,\orcidlink{0000-0003-2287-0173}} 
  \author{L.~Aggarwal\,\orcidlink{0000-0002-0909-7537}} 
  \author{H.~Aihara\,\orcidlink{0000-0002-1907-5964}} 
  \author{N.~Akopov\,\orcidlink{0000-0002-4425-2096}} 
  \author{A.~Aloisio\,\orcidlink{0000-0002-3883-6693}} 
  \author{S.~Al~Said\,\orcidlink{0000-0002-4895-3869}} 
  \author{D.~M.~Asner\,\orcidlink{0000-0002-1586-5790}} 
  \author{H.~Atmacan\,\orcidlink{0000-0003-2435-501X}} 
  \author{V.~Aushev\,\orcidlink{0000-0002-8588-5308}} 
  \author{M.~Aversano\,\orcidlink{0000-0001-9980-0953}} 
% ...
\collaboration{The Belle and Belle II Collaborations}

\begin{abstract}
We present the results of a search for the $b \to d\ell^+\ell^-$ flavor-changing neutral-current rare decays $B^{+, 0} \to (\eta, \omega, \pi^{+,0}, \rho^{+, 0}) e^+e^-$ and $B^{+, 0} \to (\eta, \omega, \pi^{0}, \rho^{+}) \mu^+\mu^-$ using a $711$~fb$^{-1}$ data sample that contains $772 \times 10^{6}$ $B\bar{B}$ events. The data were collected at the $\Upsilon(4S)$ resonance with the Belle detector at the KEKB asymmetric-energy $e^+e^-$ collider. We find no evidence for signal and set upper limits on branching fractions at the $90\%$ confidence level in the range $(3.8 - 47) \times 10^{-8}$ depending on the decay channel. The obtained limits are the world's best results. This is the first search for the channels $B^{+, 0} \to (\omega, \rho^{+,0}) e^+e^-$ and $B^{+, 0} \to (\omega, \rho^{+})\mu^+\mu^-$.
\end{abstract}

\maketitle
In the standard model (SM), the flavor-changing neutral current (FCNC) decays $B^{+, 0} \to (\eta, \omega, \pi^{+,0}, \rho^{+, 0}) \ell^+\ell^-$, $\ell = e$ or $\mu$, proceed through $b \to d\ell^+\ell^-$ transitions~\cite{chargeconjugate}. The FCNC $b \to (s, d)\ell^+\ell^-$ processes are forbidden at the tree level in the SM and proceed through loop-level diagrams. Various extensions of the SM predict the existence of new heavy particles that couple to the SM fermions and bosons. Beyond the Standard Model (BSM) physics can interfere with the SM processes, altering the physical observables and thus providing a promising avenue to search for BSM physics using rare decays~\cite{theory5, theory6}. Most experimental studies~\cite{LHCb_RK+, LHCb_RKstarplus, LHCb_RKS, babar_rk, seema, simon, LHCb_angular, simon_angularanalysis, LHCb_differential, seema_btokstll} and theoretical predictions~\cite{theory1, theory2, theory3, theory4, theory5, theory6} are focused on observables such as lepton-flavor-universality (LFU) ratios, isospin asymmetries, forward-backward asymmetries, total or differential branching fractions, angular observables, etc. in $b \to s\ell^+\ell^-$ FCNC decays. However, signatures due to BSM physics may be uniquely observed in $b \to d\ell^+\ell^-$ decays if the former is sensitive to the flavor of the quarks in the interaction~\cite{theory1}. These decays could have better sensitivity to BSM physics than $b \to s\ell^+\ell^-$ decays, as the SM branching fraction is further suppressed by a factor of $|V_{td}/V_{ts}|^2 \simeq 0.04$, where $V_{td}$ and $V_{ts}$ are elements of the Cabibbo-Kobayashi-Maskawa quark-mixing matrix~\cite{ckm1, ckm2}. Typical branching fractions of $b \to d\ell^+\ell^-$ decays are of ${\cal O}(10^{-8})$ or smaller in the SM~\cite{btodll_theory1, btodll_theory2, btodll_theory3}, making them a challenging target for experiments. 

Previously, BaBar, Belle, and LHCb have searched for several of these decay processes. The best upper limits (ULs) on the $B^0 \to \eta\ell^+\ell^-$ and $B^0 \to \pi^0\ell^+\ell^-$ branching fractions are from BaBar~\cite{babar1} using a $428$~fb$^{-1}$ data sample, while the best $B^+ \to \pi^+e^+e^-$ branching fraction UL is from a 605~fb$^{-1}$ data sample of Belle~\cite{belle1}. LHCb~\cite{LHCb1, pipimumu} has measured the branching fractions for the decays $B^+ \to \pi^+\mu^+\mu^-$, $B^0 \to \pi^+\pi^-\mu^+\mu^-$, and $B^0 \to \rho^0\mu^+\mu^-$ to be $(1.78 \pm 0.23) \times 10^{-8}$, $(2.11 \pm 0.52) \times 10^{-8}$, and $(1.98 \pm 0.53) \times 10^{-8}$, respectively, using $3$~fb$^{-1}$ of data. We perform searches for rare $B^{+, 0} \to (\eta, \omega, \pi^{+,0}, \rho^{+, 0}) \ell^+\ell^-$ decays using a 711~fb$^{-1}$ data sample that contains $772 \times 10^6$ $B\bar{B}$ events, collected at the $\Upsilon(4S)$ resonance with the Belle detector at the KEKB asymmetric-energy $e^+e^-$ collider. The data are converted into the Belle II analysis software framework (BASF2) ~\cite{basf2} format using the B2BII software package~\cite{b2bii}. Searches are conducted in both the electron and muon channels. Tests of LFU in $b \to d\ell\ell$ transitions are possible if these modes are seen in the data~\cite{LFU_btodll}. 

The Belle detector~\cite{belle_detector} is a large-solid-angle magnetic spectrometer. The inner part is composed of a silicon vertex detector (SVD), a 50-layer central drift chamber (CDC), an array of aerogel threshold Cherenkov counters (ACC), a barrel-like arrangement of time-of-flight scintillation counters (TOF), and an electromagnetic calorimeter (ECL) comprised of CsI(Tl) crystals. All these sub-detectors are located inside a superconducting solenoid coil that provides a $1.5$~T magnetic field. An iron flux-return yoke placed outside the coil is instrumented with resistive plate chambers to detect $K_{L}^{0}$ mesons and muons (KLM). Two inner detector configurations are used: a $2.0$ cm radius beam pipe and a three-layer SVD for the first sample of $140$ fb$^{-1}$; and a $1.5$ cm radius beam pipe, a four-layer SVD, and a small-inner-cell CDC for the remaining $571$~fb$^{-1}$~\cite{svd2}.

We use Monte Carlo (MC) simulated events to study the properties of signal decays and to suppress various background sources. The  $B^{+, 0} \to (\eta, \omega, \pi^{+,0}, \rho^{+, 0}) \ell^+\ell^-$ decays are generated with the EvtGen package~\cite{evt_gen} using the BTOSLLBALL~\cite{BTOSLLBALL} decay model. The PHOTOS package~\cite{photos} is used to incorporate final-state radiation effects, while GEANT3~\cite{geant3} is used for detector simulation. We study the expected background contributions using simulated samples corresponding to an integrated luminosity six times that of the Belle data sample. The background sample includes on-resonance $\Upsilon(4S) \to B\bar{B}$ (other $B$ decay) and continuum $e^+e^- \to q\bar{q}$ events with $q \in {u, d, s, c}$, which are generated using the EvtGen, PYTHIA~\cite{pythia}, and PHOTOS packages with interference effects due to final state radiation switched on for the latter.

We reconstruct  $B^0 \to \eta e^+e^-$, $B^0 \to \eta \mu^+\mu^-$, $B^0 \to \omega e^+e^-$, $B^0 \to \omega\mu^+\mu^-$, $B^0 \to \pi^0e^+e^-$, $B^0 \to \pi^0\mu^+\mu^-$, $B^+ \to \pi^+e^+e^-$, $B^0 \to \rho^0e^+e^-$, $B^+ \to \rho^+e^+e^-$, and $B^+ \to \rho^+\mu^+\mu^-$ decays. The charged particles $\pi^{\pm}$, $\mu^{\pm}$, and $e^{\pm}$ are selected to originate from the interaction point by requiring their impact parameters to be less than $4.0$~cm along the $z$~axis (direction opposite to the $e^+$ beam), and less than $1.0$~cm in the transverse plane. We apply a minimum of $100$~MeV$/c$ on their transverse momentum to reduce the background from low-momentum particles. The muon candidates are selected using a likelihood ratio ${\mathcal R}_{\mu} = {\mathcal L}^{}_\mu / ({\mathcal L}^{}_\mu + {\mathcal L}^{}_{\pi} + {\mathcal L}^{}_{K})$, where ${\mathcal L}^{}_{\mu}$, ${\mathcal L}^{}_{\pi}$, and ${\mathcal L}^{}_{K}$ are the likelihood values obtained for the muon, pion, and kaon hypotheses, respectively, based on information from the KLM. The muon candidates are required to have a minimum momentum of $0.8$ GeV$/c$ to ensure they reach the KLM. We apply ${\mathcal R}_{\mu}>0.9$, corresponding to an efficiency of $89\%$ with a pion (or kaon) misidentification rate of approximately $1.5\%$~\cite{muon_identification1}. The electron candidates are required to have a minimum momentum of $0.5$~GeV$/c$ and an electron likelihood ratio ${\mathcal R}_{e} = {\mathcal L}^{}_e / ({\mathcal L}^{}_e + {\mathcal L}_{\slashed{e}})>0.9$, where ${\mathcal L}^{}_e$ and ${\mathcal L}_{\slashed{e}}$ are the likelihood values for electron and non-electron hypotheses, respectively. These likelihoods are calculated with the ratio of calorimetric cluster energy to the track momentum, the shower shape in the ECL, the matching of the track with the ECL cluster, the specific ionization in the CDC, and the number of photoelectrons in the ACC~\cite{electron_identification}. The electron selection efficiency is $92\%$ with a pion misidentification rate of less than $1\%$. The photon candidates are identified from energy clusters in the ECL that are not associated with any charged track. The photon energy is required to be greater than 50 MeV if reconstructed in the barrel, and greater than 100 or 150 MeV if reconstructed in the forward or backward endcap regions, respectively, to remove beam-induced background. The forward endcap, barrel, and backward endcap regions of the ECL are given by $12^{\circ} < \theta < 31^{\circ}$, $32^{\circ} < \theta < 129^{\circ}$, and $132^{\circ} < \theta < 157^{\circ}$, respectively, where $\theta$ is the polar angle in the laboratory frame with respect to the $z$--axis. The ratio of the energy deposited in a $3\times 3$ array of crystals centered on the crystal with the highest energy to the energy deposited in the corresponding $5 \times 5$ array is required to be greater than 0.80 to reject showers produced by hadrons. The energy loss due to bremsstrahlung for the electron candidate is recovered by considering the energy of all photons found in a 50~mrad cone along its initial momentum direction. The pion candidates are selected using a likelihood ratio ${\mathcal R}^{}_{\pi/K} = {\mathcal L}^{}_\pi / ({\mathcal L}^{}_\pi + {\mathcal L}^{}_K )$. Each candidate's likelihood is calculated based on the number of photoelectrons in the ACC, the specific ionization in the CDC, and the flight time in the TOF. The requirement ${\mathcal R}^{}_{\pi/K}>0.6$ is $89\%$ efficient for pions, and has a misidentification rate of $\sim 8\%$ for kaons~\cite{hadron_identification}. Candidate $\pi^0 \to \gamma\gamma$ decays are reconstructed from photon pairs that have an invariant mass satisfying $124 < M_{\gamma\gamma}< 145$~MeV$/c^{2}$, this region corresponds to about $\pm 3\sigma$ of the invariant mass resolution around the nominal $\pi^0$ mass~\cite{PDG}. A mass-constrained fit is subsequently performed to improve the $\pi^0$ momentum resolution. We apply a minimum momentum requirement of $1$~GeV$/c$ on the pion candidates only for the decays $B^{+, 0} \to \pi^{+, 0}e^+e^-$ to suppress background from low-multiplicity two-photon processes. 

The $\eta$ meson candidates are reconstructed in the decay modes $2\gamma$ and $\pi^+\pi^-\pi^0$ in the mass range $M_{\eta} \in [530, 559]$~MeV$/c^{2}$. Similarly, the $\omega$ meson candidates are reconstructed in the decay mode $\pi^+\pi^-\pi^0$ in the mass range $M_{\omega} \in [768, 795]$~MeV$/c^{2}$. The selected mass windows for $\eta$ and $\omega$ are within $\pm 2\sigma$ of the known masses~\cite{PDG}. The $\pi^+$ candidates are combined with $\pi^-$ or $\pi^0$ candidates to form $\rho^0$ or $\rho^+$ meson in the mass range $M_{\rho^{+, 0}} \in [650, 900]$~MeV$/c^{2}$. The charged (neutral) $B$ candidates are reconstructed by combining light meson candidates with selected $e^+e^-$ or $\mu^+\mu^-$ candidates. The kinematic variables that distinguish signal from background are the beam-energy constrained mass $M_{\rm{bc}}$ and the energy difference $\Delta E$,

\begin{align*}
    M_{\rm{bc}} &= \sqrt{(E^{\ast}_{\rm beam}/c^{2})^{2} - (p^{\ast}_{B}/c)^{2}},  \hspace{1.5cm} (1)\\
    \Delta E &=  E^{\ast}_{B} - E^{\ast}_{\rm beam}, \hspace{3.45cm} (2)
\end{align*}
where $E^{\ast}_{\rm beam}$ is the beam energy, while $E^{\ast}_{B}$ and $p^{\ast}_{B}$ are the energy and momentum of the $B$ candidate, respectively. These quantities are calculated in the $e^{+}e^{-}$ center-of-mass (c.m.)\ frame. We retain the candidates that satisfy $M_{\rm{bc}}>5.2$~GeV$/c^2$ and $-0.15<\Delta E<0.10$~GeV for further analysis. For signal events, $M_{\rm{bc}}$ and $\Delta E$ should peak at the $B$ mass and zero, respectively. The mass windows for $\eta$, $\omega$, and $\rho^{0,+}$ mesons are obtained by maximizing the figure of merit (FOM) = ${\varepsilon}/({\mathit{a}/2 + \sqrt{B}})$~\cite{FOM}, where $\varepsilon$ and $B$ are the signal efficiency in MC simulated events and the number of background events in the signal region, $M_{\rm{bc}}>5.27$~GeV$/c^{2}$ and $|\Delta E|<0.05$~GeV. The $\mathit{a}$ term denotes the significance, where we use $\mathit{a} = 3\sigma$.

The contributions from the charmonium decays $B^{+, 0} \to (\eta, \omega, \pi^{+,0}, \rho^{+, 0}) J/\psi(\to \ell^+\ell^-)$ and $(\eta, \omega, \pi^{+,0}, \rho^{+, 0}) \psi(2S) (\to \ell^+\ell^-)$, are suppressed by rejecting events with invariant-mass squared of the lepton pair in the ranges $8.5~(8.8) < q^2 < 10.22~(9.9)$~GeV$^2/c^4$ and $13~(13)<q^2<14~(14)$~GeV$^2/c^4$, respectively, for the electron (muon) channels. An additional rejection criterion $q^2<0.045$~GeV$^2/c^4$ is applied to suppress possible contamination from converted photons or $\pi^0$ Dalitz decays. 

We find significant contributions from continuum processes and other $B$ decays after signal reconstruction. The continuum background typically has a back-to-back topology, in contrast to $B\bar{B}$ events which are produced almost at rest in the c.m.\ frame resulting in a more spherical topology. To reduce the background contribution, we use a boosted decision tree (BDT)~\cite{fastBDT} technique. The variables used are, the ratio of the second to zeroth Fox-Wolfram moment~\cite{R22}; the cosine of the angle between the $B$ flight direction and the $z$--axis; the cosine of the angle between the thrust axis of the $B$ candidate and that of the rest of the event (ROE), $i.e.,$ not associated to the signal candidate, in the c.m.\ frame; the magnitude of the signal $B$ thrust; the magnitude of the ROE thrust~\cite{ROE_thrust}; the difference between the $z$ coordinate of the decay vertices of the signal $B$ and the ROE; the separation between the two lepton tracks along the $z$--axis; the azimuthal separation between two leptons; the separation between the light meson and high-momentum lepton track along the $z$--axis; the $B$ vertex fit probability; the sum of energy of the tracks and clusters of the ROE; the missing mass squared over missing energy of the ROE; and the output of a BDT-based flavor tagger~\cite{qr}. We determine an optimal requirement on the BDT output for each decay channel by maximizing the FOM. These BDT requirements reduce the background by $93-98\%$ with $25-55\%$ signal efficiency loss, depending on the decay channel.

With the above selection criteria applied, the average candidate multiplicity per event is found to be $1.03-1.61$ from signal MC samples across the decay channels. When there are multiple candidates, we select the candidate having the smallest $\chi^2$ value from the $B$ decay vertex fit. This procedure selects the correct signal candidate $70 - 90\%$ of the time, depending on the decay channel. 

We perform a two-dimensional unbinned extended maximum-likelihood fit to the $M_{\rm{bc}}$ and $\Delta E$ distributions to extract the signal yields. In $M_{\rm{bc}}$ the signal component is modeled with a Crystal Ball~\cite{CB} or a Crystal Ball with a Gaussian and in $\Delta E$ with Johnson's $S_U$~\cite{johnson} or Johnson's $S_U$ with a Gaussian. The signal parameters of the fit to data are fixed to those in simulation while the width and mean are modified with width scale factors and mean shifts obtained from $B^{+, 0} \to J/\psi( \to \ell^+\ell^-)\pi^{+, 0}$ control samples, to take into account the differences between data and simulation. The combinatorial backgrounds are modeled with an ARGUS~\cite{ARGUS} shape and a polynomial of first or second order for $M_{\rm{bc}}$ and $\Delta E$, respectively. The background shape parameters are floated in the fit in addition to their yield. We also look for background from specific decay channels that can mimic the signal and peak in the $M_{\rm{bc}}$ or $\Delta E$ signal region, or both using MC simulation. For the $B^+ \to \pi^+e^+e^-$ channel, the background from $B^+ \to K^+e^+e^-$ decays peaks in the $M_{\rm{bc}}$ signal region but is shifted to negative values of $\Delta E$. This background is studied using a MC sample of that decay. The peaking background in $M_{\rm{bc}}$ and $\Delta E$ are both fit with a combination of Gaussian and asymmetric Gaussian probability density functions (PDFs). For the $B^0 \to \rho^0 e^+e^-$ and $B^+ \to \rho^+\mu^+\mu^-$ decay channels a peaking background arises when one of the two leptons originating from a $J/\psi$ is combined with a kaon from a $K^{\ast}$ resonance ($K^{\ast 0}$, $K_{0}^{\ast 0}$, $K_{2}^{\ast 0}$, etc) that is misidentified as a lepton. This background peaks in the $M_{\rm{bc}}$ signal region but is shifted towards negative values of $\Delta E$. This contribution is studied using an inclusive $J/\psi$ MC sample with an integrated luminosity corresponding to $100$ times the Belle data sample. The contribution from this peaking background for both $M_{\rm{bc}}$ and $\Delta E$ is fit with a combination of Gaussian and asymmetric Gaussian PDFs. The PDF shape parameters are fixed in the signal yield extraction procedure while the yield is floated. For the $B^+ \to \rho^+\mu^+\mu^-$ channel, we find an additional peaking contribution from $B^+ \to \rho^+\bar{D}{}^0(\to K^+\pi^-)$, where both the $K^+$ and $\pi^-$ are misidentified as muon candidates. We also apply a veto on the mass of the muon pair, $M_{\mu^+\mu^-} \notin [1.858-1.881]$~GeV$/c^{2}$, to reduce this background, where the invariant mass is calculated by reassigning the faking particle mass hypotheses to the muon candidates. We find background from charmless decays, $i.e.,$ $B^0 \to \eta K^+\pi^-$ contributions to the $B^0 \to \eta \mu^+\mu^-$ channel, and $B^0 \to \pi^0K^+\pi^-$ and $B^0 \to \pi^0K^+K^-$ contributions to the $B^0 \to \pi^0\mu^+\mu^-$ channel, and the PDF shapes for these peaking contributions are determined with simulation and corresponding yields are floated. 
\begin{figure} []
\begin{center}
\begin{overpic}[width=1\columnwidth]{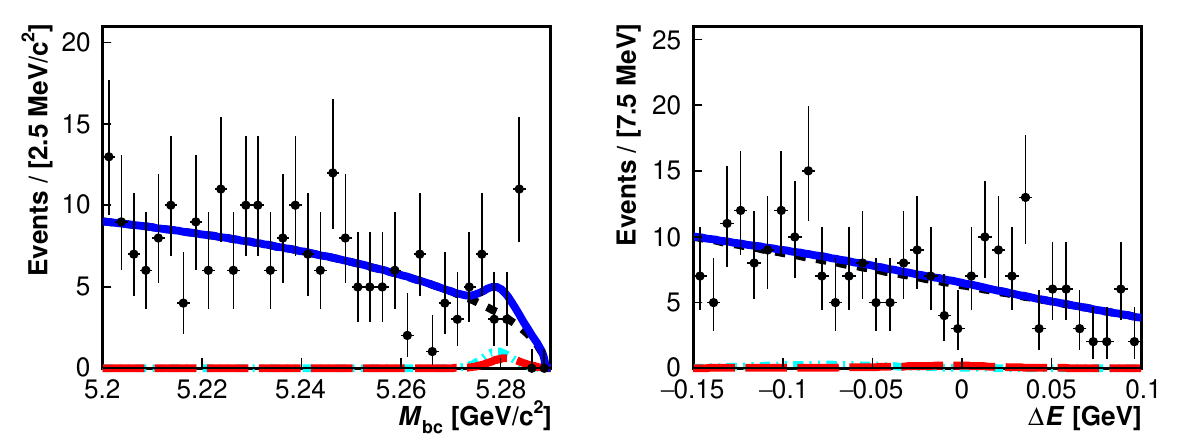}
 \put (8.5,36) {\scriptsize{{{\textit {\textbf {Belle}}} }}}
 \put (15,31) {(a) $B^0 \to \eta \ell^+\ell^-$}
 \put (58.5,36) {\scriptsize{{{\textit {\textbf {Belle}}} }}}
 \put (65,31) {(a) $B^0 \to \eta \ell^+\ell^-$}
 \end{overpic}
  \begin{overpic}[width=1\columnwidth]{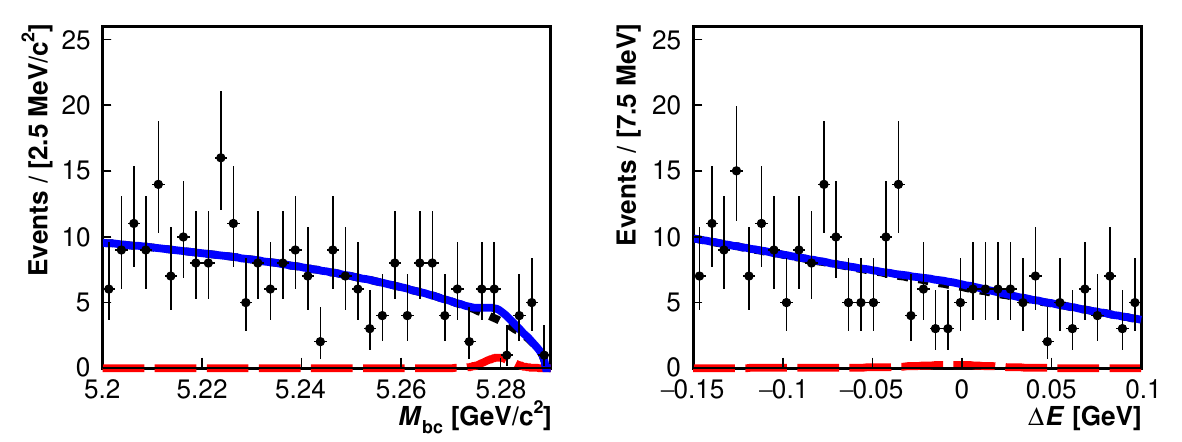}
  \put (8.5,36) {\scriptsize{{{\textit {\textbf {Belle}}} }}}
 \put (15,31) {(b) $B^0 \to \omega \ell^+\ell^-$}
 \put (58.5,36) {\scriptsize{{{\textit {\textbf {Belle}}} }}}
 \put (65,31) {(b) $B^0 \to \omega \ell^+\ell^-$}
\end{overpic}
\begin{overpic}[width=1\columnwidth]{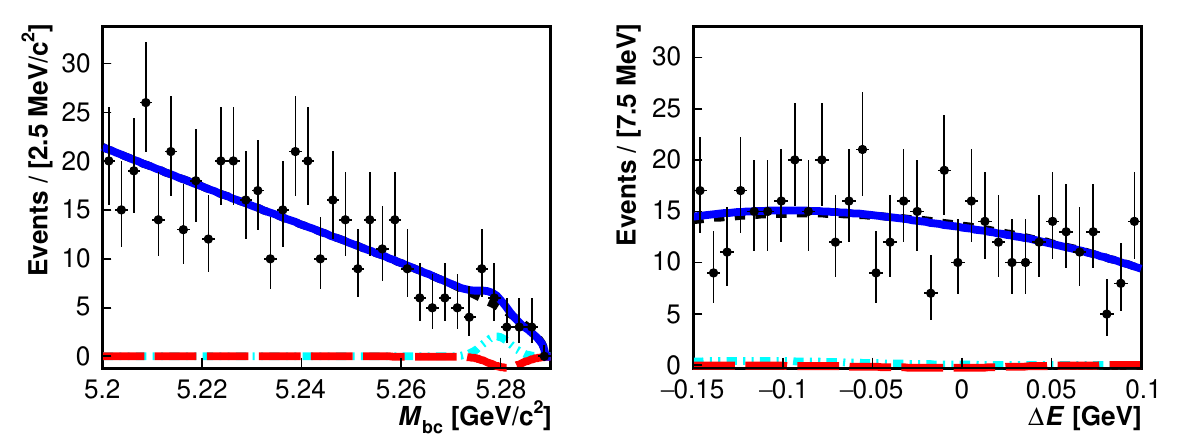}
 \put (8.5,36) {\scriptsize{{{\textit {\textbf {Belle}}} }}}
 \put (15,31) {(c) $B^0 \to \pi^0 \ell^+\ell^-$}
 \put (58.5,36) {\scriptsize{{{\textit {\textbf {Belle}}} }}}
 \put (65,31) {(c) $B^0 \to \pi^0 \ell^+\ell^-$}
\end{overpic}
\begin{overpic}[width=1\columnwidth]{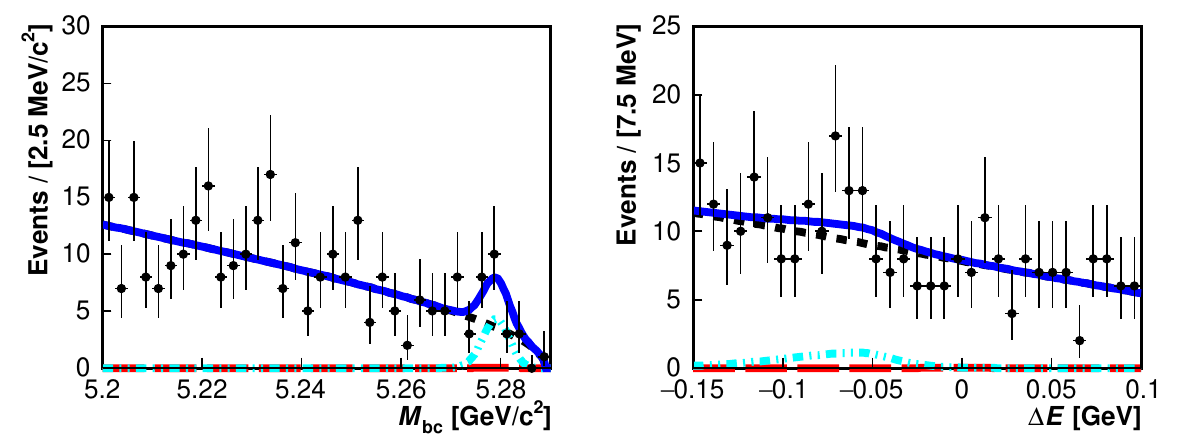}
\put (8.5,36) {\scriptsize{{{\textit {\textbf {Belle}}} }}}
 \put (15,31) {(d) $B^+ \to \pi^+ e^+e^-$}
 \put (58.5,36) {\scriptsize{{{\textit {\textbf {Belle}}} }}}
 \put (65,31) {(d) $B^+ \to \pi^+ e^+e^-$}
\end{overpic}
\begin{overpic}[width=1\columnwidth]{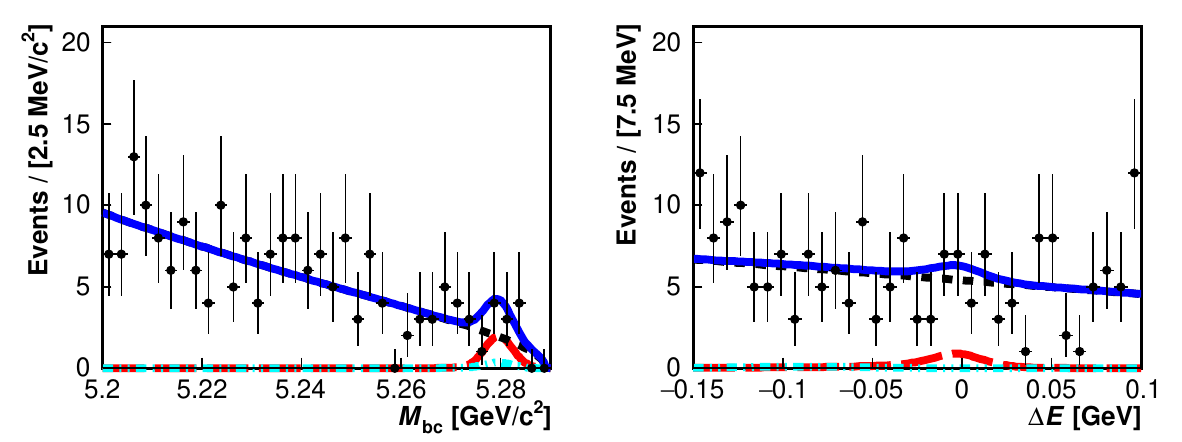}
\put (8.5,36) {\scriptsize{{{\textit {\textbf {Belle}}} }}}
 \put (15,31) {(e) $B^0 \to \rho^0 e^+e^-$}
 \put (58.5,36) {\scriptsize{{{\textit {\textbf {Belle}}} }}}
 \put (65,31) {(e) $B^0 \to \rho^0 e^+e^-$}
\end{overpic}
\begin{overpic}[width=1\columnwidth]{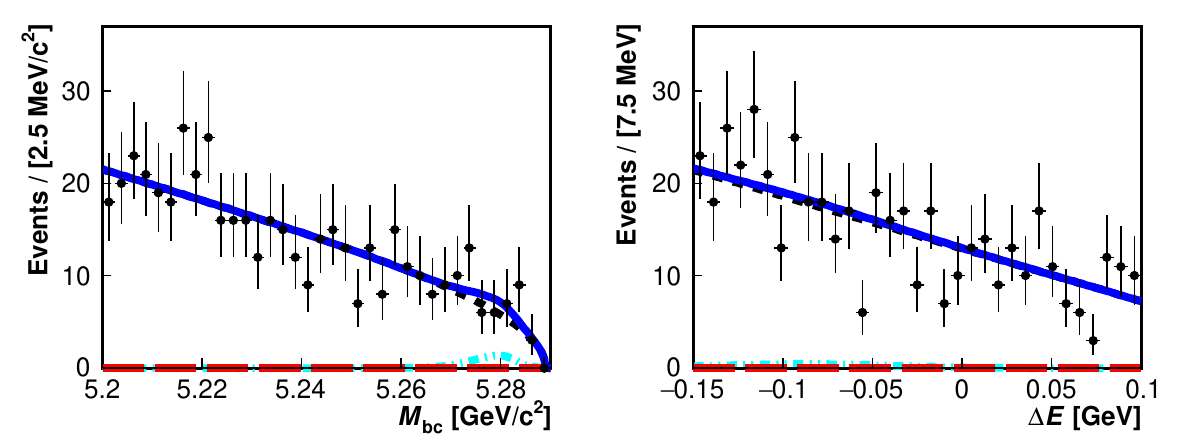}
\put (8.5,36) {\scriptsize{{{\textit {\textbf {Belle}}} }}}
 \put (15,31) {(f) $B^+ \to \rho^+ \ell^+\ell^-$}
 \put (58.5,36) {\scriptsize{{{\textit {\textbf {Belle}}} }}}
 \put (65,31) {(f) $B^+ \to \rho^+ \ell^+\ell^-$}
\end{overpic}
\end{center}
  \caption{$M_{\rm{bc}}$~(left) and $\Delta E$~(right) distributions of two-dimensional unbinned extended maximum-likelihood fits to data after signal selection for (a) $B^{0} \to \eta \ell^+\ell^-$ (first-row), (b) $B^0 \to \omega \ell^+\ell^-$ (second-row), (c) $B^0 \to \pi^0 \ell^+\ell^-$ (third-row), (d) $B^+ \to \pi^+e^+e^-$ (fourth-row), (e) $B^0 \to \rho^0 e^+e^-$ (fifth-row), and (f) $B^+ \to \rho^+\ell^+\ell^-$ (sixth-row). Points with error bars are the data; blue solid curves are the fitted results for the signal-plus-background hypothesis; red long-dashed curves denote the signal component; black dashed and cyan dash-dotted curves are combinatorial and peaking backgrounds, respectively.}
  \label{dataFigure}
\end{figure}
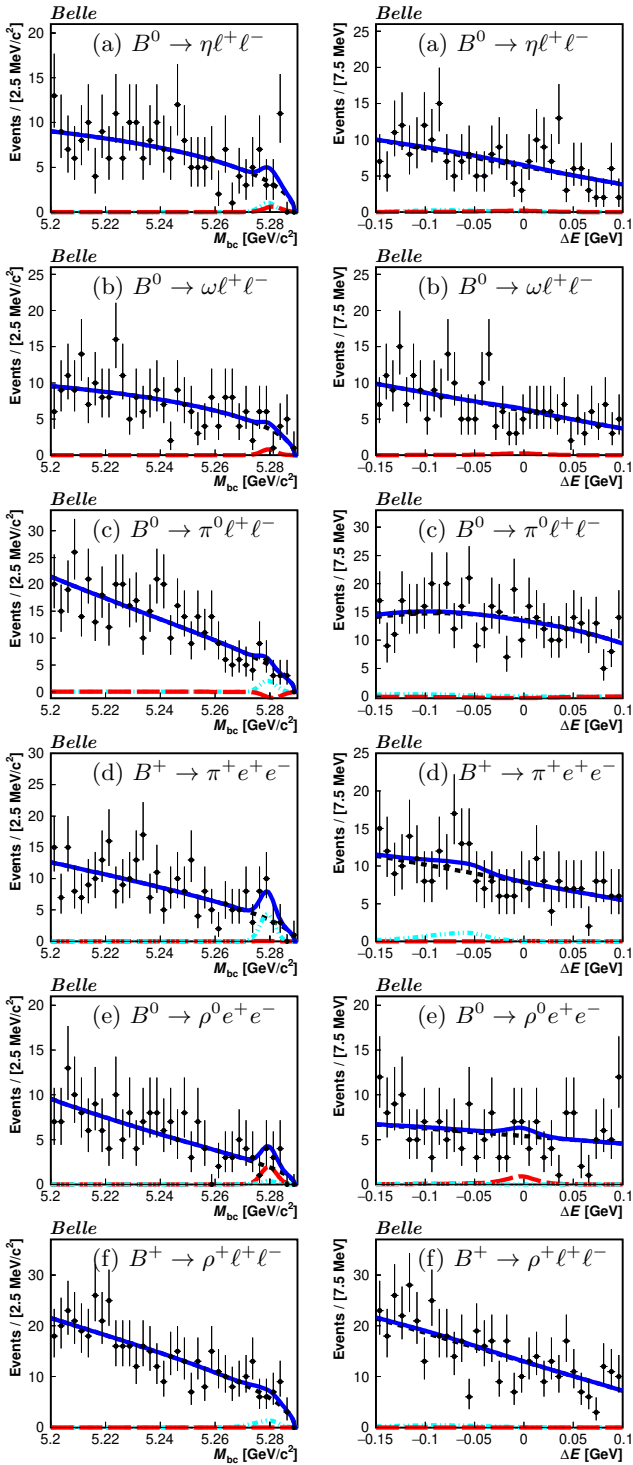

The fits are performed separately for each of the charged and neutral $B$ decay channels containing $\mu^+\mu^-$ and $e^+e^-$ in the final state. To extract the combined signal yield for modes containing $\ell^+\ell^-$ ($e^+e^-$ and $\mu^+\mu^-$) in the final state, we perform a simultaneous fit of the $e^+e^-$ and $\mu^+\mu^-$ channels, where the background shape parameters are common between the two samples. The fit results are shown in Fig.~\ref{dataFigure}. There is no significant excess of signal found in any of the decay channels.

We calculate the ULs at the $90\%$ confidence level (CL) for these decay channels using a frequentist approach. In this method, 10,000 MC experiments are generated using signal and background PDFs considering different numbers of signal events, $N_{\rm{sig}}(\rm{gen})$, and background events obtained from fitting. The simulated data sets are fit, and we calculate the fraction of MC experiments that have fitted yields less than or equal to that observed in data, $i.e.,$ $N_{\rm{sig}} \leq N_{\rm{sig}} (\rm{data})$. For channels with negative signal yield, the $N_{\rm{sig}} (\rm{data})$ is assumed to be 0 for UL calculation. The $90\%$ CL UL is the value of $N_{\rm{sig}} (\rm{gen})$ for which $10\%$ of the experiments have $N_{\rm{sig}}\leq N_{\rm{sig}}(\rm{data})$, defined as $N^{\rm{UL}}_{\rm{sig}}$. 

Several sources of systematic uncertainties contribute to the branching fraction UL measurement. The systematic uncertainty due to $\pi^+$ identification is $0.5\%$ from a study of a $D^{\ast +} \to D^0(\to K^-\pi^+)\pi^+$ sample. The $\pi^0$ efficiency is studied using $\tau^- \to \pi^-\pi^0\nu_{\tau}$ decays and found to have a systematic uncertainty of $2.3\%$. The systematic uncertainties arising due to lepton identification are $0.3\%$ and $0.4\%$ for each selected muon and electron, respectively, calculated from an inclusive $J/\psi$ sample. The photon detection uncertainty is $2.0\%$ determined from radiative Bhabha and $B^0 \to K^{\ast 0}\gamma$ samples. The systematic uncertainty due to charged track reconstruction is $0.35\%$ per track estimated by using partially reconstructed $D^{\ast +} \rightarrow D{}^{0}\pi^+$, $D{}^{0}\rightarrow \pi^-\pi^+K_{S}^{0}$, and $K_{S}^{0} \rightarrow \pi^+\pi^-$ events. The signal decay model systematic uncertainty is obtained by replacing the BTOSLLBALL decay model with LCSR~\cite{LCSR} and QUARK~\cite{QUARK} models, and the maximum deviation of the signal efficiencies from the nominal is assigned as a systematic uncertainty, which has an effect of less than $1\%$. The uncertainty in efficiency due to the limited MC sample size is less than $1\%$. To account for the uncertainty arising from the mass window requirements for $\rho$, $\eta$, and $\omega$ mesons, the change in signal yield in MC by varying the window to data resolution is assigned as a systematic. This varies between $1$ and $3\%$. To assess the potential bias due to the method of best candidate selection, this is changed to random candidate selection, and the difference in branching fraction ULs at the $90\%$ confidence level has a systematic uncertainty of less than $1\%$. The effect of the BDT, used for background suppression, is studied using $B^{+, 0}\to J/\psi(\to \ell^+\ell^-)\pi^{+, 0}$ channels by taking the ratio in efficiencies between the data and simulation, which results in systematic uncertainties of $1-7\%$, depending on the decay channel. The uncertainty in the number of $B\bar{B}$ events is $1.4\%$. The systematic uncertainty in both ${\cal B}[\Upsilon(4S) \to B^+B^-]$ and ${\cal B}[\Upsilon(4S) \to B^0\bar{B}{}^0]$ is $2.4\%$~\cite{seema_fpm00}. The shape parameters fixed in the fit are varied by $\pm 1\sigma$, determined from $10^6$ generated signal MC events or a $B^{+, 0}\to J/\psi(\to \ell^+\ell^-)\pi^{+, 0}$ control sample, from their mean values and the deviation from the nominal fit value of $N_{\rm{sig}}$ is the uncertainty due to the signal and background shapes: this is found to be less than $1\%$. 

The branching fractions ULs are calculated using the formula
\begin{align*}
     {\cal B}^{\rm{UL}} &= \dfrac{N^{\rm{UL}}_{\rm{sig}}}{2 f^{\pm(00)} N_{B\bar{B}} \varepsilon}. \hspace{1.5cm} (3)
\end{align*}
Here, $f^{\pm(00)}$ is the branching fraction ${\cal B}[\Upsilon(4S) \to B^{+}B^{-}]$ = $(51.6 \pm 1.2)\%$  $({\cal B}[\Upsilon(4S) \to B^{0}\bar{B}{}^0] = (48.4 \pm 1.2) \%)$ for charged (neutral) $B$ decays~\cite{seema_fpm00}; $N_{B\bar{B}}$ and $\varepsilon$ are the number of $B\bar{B}$ events~=~$(772 \pm 11) \times 10^6$ and data-MC difference corrected signal MC efficiency, respectively. The systematic uncertainties in ${\cal B}^{\rm{UL}}$ are included by smearing $N_{\rm{sig}}$ with the fractional systematic uncertainties described above. The results are listed in Table~\ref{results}.
\begin{table*} [htbp]
\caption{${\cal B}^{\rm{UL}}$ for $b \to de^+e^-$, $b \to d\mu^+\mu^-$, and $b \to d\ell^+\ell^-$ decays. The columns correspond to decay channels, signal yields ($N_{\rm{sig}}$), $90\%$ CL signal yield upper limits ($N^{\rm{UL}}_{\rm{sig}}$), data-MC difference corrected signal MC efficiencies ($\varepsilon$), branching fraction $90\%$ CL upper limits ($\cal B^{\rm{UL}}$), previous branching fraction $90\%$ CL upper limits (Previous $\cal B^{\rm{UL}}$), branching fractions ($\cal B$), and branching fraction theoretical predictions (Theory $\cal B$). }
\label{results}
\begin{center}
\begin{tabular*}{0.95\linewidth}{@{\extracolsep{\fill}}l cccccccc} 
\hline \hline 
\\ & & & & & Previous & & Theory & \\ \vspace{-0.35cm}
\\ Channel & $N_{\rm{sig}}$ & $N^{\rm{UL}}_{\rm{sig}}$ & $\varepsilon~(\%)$ & \phantom{0}$\cal B^{\rm{UL}}$ ($10^{-8}$) & $\cal B^{\rm{UL}}$ ($10^{-8}$) & $\cal B$ ($10^{-8}$) & $\cal B$ ($10^{-8}$) \\ \\ \hline \\
$B^0 \to \eta e^+e^-$ & $\phantom{-}0.0^{+1.4}_{-1.0}$ & \phantom{0}3.1& $\phantom{0}3.9 $ &  $<10.5$ & $<10.8$~\cite{babar1} & $\phantom{+}\phantom{0}0.0^{+4.9}_{-3.4} \pm 0.1$ & $-$\\ 
$B^0 \to \eta \mu^+\mu^-$ & $\phantom{-}0.8^{+1.5}_{-1.1}$ & \phantom{0}4.2 & $\phantom{0}5.9 $ & $<\phantom{0}9.4$ & $<11.2$~\cite{babar1}& $\phantom{+}\phantom{0}1.9^{+3.4}_{-2.5} \pm 0.2$ & $-$\\ 
$B^0 \to \eta \ell^+\ell^-$ & $\phantom{-}0.5^{+1.0}_{-0.8}$&  \phantom{0}1.8& $\phantom{0}4.9 $ &  $<\phantom{0}4.8$ & $<\phantom{0}6.4$~\cite{babar1} & $\phantom{+}\phantom{0}1.3^{+2.8}_{-2.2} \pm 0.1$ & $-$\\ \\ 
$B^0 \to \omega e^+e^-$ & $-0.3^{+3.2}_{-2.5}$&\phantom{0}3.7 & $\phantom{0}1.6$ & $<30.7$ & $-$ & $\phantom{0}-2.1^{+26.5}_{-20.8} \pm 0.2$ & $-$\\ 
$B^0 \to \omega \mu^+\mu^-$ &$\phantom{-}1.7^{+2.3}_{-1.6}$& \phantom{0}5.5& $\phantom{0}2.9$ & $<24.9$ & $-$ & $\phantom{-}\phantom{0}\phantom{0}7.7^{+10.8}_{-\phantom{0}7.5} \pm 0.6$ & $-$\\ 
$B^0 \to \omega \ell^+\ell^-$ & $\phantom{-}1.0^{+1.8}_{-1.3}$& \phantom{0}3.6& $\phantom{0}2.2$ & $<22.0$ & $-$& $\phantom{+}\phantom{0}\phantom{0}6.4^{+10.7}_{-\phantom{0}7.8} \pm 0.5$ & $-$\\ \\ 
$B^0 \to \pi^0 e^+e^-$ &$-2.9^{+1.8}_{-1.4}$ & \phantom{0}4.0&$\phantom{0}6.7$ & $<\phantom{0}7.9$ & $<\phantom{0}8.4$~\cite{babar1} & $\phantom{0}-5.8^{+3.6}_{-2.8} \pm 0.5$ &  $-$\\ 
$B^0 \to \pi^0 \mu^+\mu^-$ &$-0.5^{+3.6}_{-2.7}$ & \phantom{0}6.1&$13.7$ & $<\phantom{0}5.9$ & $<\phantom{0}6.9$~\cite{babar1} & $\phantom{0}-0.4^{+3.5}_{-2.6} \pm 0.1$ &  $-$\\ 
$B^0 \to \pi^0 \ell^+\ell^-$ &$-1.8^{+1.6}_{-1.1}$ & \phantom{0}2.9& $10.2$&  $<\phantom{0}3.8$ & $<\phantom{0}5.3$~\cite{babar1} & $\phantom{0}-2.3^{+2.1}_{-1.5} \pm 0.2$ & $0.91^{+0.34}_{-0.29}$~\cite{btodll_theory3}\\  \\ 
$B^+ \to \pi^+ e^+e^-$ &$\phantom{-}0.1^{+2.5}_{-1.6}$& \phantom{0}5.0& $11.5$ & $<\phantom{0}5.4$ & $<\phantom{0}8.0$~\cite{belle1} & $\phantom{+}\phantom{0}\phantom{0}0.1^{+2.7}_{-1.8} \pm 0.1$ &  $1.96 \pm 0.21$~\cite{btodll_theory2}\\ \\
$B^0 \to \rho^0 e^+e^-$ &$\phantom{-}5.6^{+3.5}_{-2.7}$& 10.8& $\phantom{0}3.2$ & $<45.5$ & $-$  & $\phantom{-}23.6^{+14.6}_{-11.2} \pm 1.1$ & $-$\\  \\
$B^+ \to \rho^+ e^+e^-$ &$-4.4^{+2.3}_{-2.0}$ & \phantom{0}5.3& $\phantom{0}1.4$ & $<46.7$ & $-$ & $-38.2^{+24.5}_{-17.2} \pm 3.4$ &  $4.20^{+0.88}_{-0.78}$~\cite{btodll_theory2}\\ 
$B^+ \to \rho^+ \mu^+\mu^-$ & $\phantom{-}3.0^{+4.0}_{-3.0}$& \phantom{0}8.7& $\phantom{0}2.9$ & $<38.1$ & $-$ & $\phantom{+}13.0^{+17.5}_{-13.3} \pm 1.1$ &  $4.03^{+0.83}_{-0.75}$~\cite{btodll_theory2}\\ 
$B^+ \to \rho^+ \ell^+\ell^-$ &$\phantom{-}0.4^{+2.3}_{-1.8}$& \phantom{0}3.0& $\phantom{0}2.0$ & $<18.9$  & $-$  &  $\phantom{-}\phantom{0}2.5^{+14.6}_{-11.8} \pm 0.2$ &  $-$ \\ \\ \hline \hline
\end{tabular*}
\end{center}
\end{table*} 

In summary, we have searched for the rare decays $B^{+, 0} \to (\eta, \omega, \pi^{+,0}, \rho^{+, 0}) e^+e^-$ and $B^{+, 0} \to (\eta, \omega, \pi^{0}, \rho^{+}) \mu^+\mu^-$, which involve $b \to d \ell^+\ell^-$ transitions, using a $711$~fb$^{-1}$ data sample of Belle. We find no evidence for the signal in any of the decay channels and set $90\%$ confidence-level upper limits on the branching fractions in the range $(3.8 - 47) \times 10^{-8}$, depending on the decay channel. The world’s best limits are obtained. This is the first search for the channels $B^{+, 0} \to (\omega, \rho^{+,0}) e^+e^-$ and $B^{+, 0} \to (\omega, \rho^{+})\mu^+\mu^-$. Our branching fraction results for $B^+ \to \pi^+e^+e^-$ and $B^0 \to \rho^0 e^+ e^-$, though statistically limited, are consistent with measurements of the $B^+ \to \pi^+\mu^+\mu^-$ and $B^0 \to \rho^0\mu^+\mu^-$ branching fractions from LHCb~\cite{LHCb1, pipimumu} within $1-2\sigma$. These results are consistent with lepton-flavor-universality in $b \to d\ell^+\ell^-$ transitions. These results will constrain the development of BSM models for $b \to d\ell^+\ell^-$ transitions.

\section{Acknowledgments}
This work, based on data collected using the Belle detector, which was
operated until June 2010, was supported by 
the Ministry of Education, Culture, Sports, Science, and
Technology (MEXT) of Japan, the Japan Society for the 
Promotion of Science (JSPS), and the Tau-Lepton Physics 
Research Center of Nagoya University; 
the Australian Research Council including grants
DP210101900, 
DP210102831, 
DE220100462, 
LE210100098, 
LE230100085; 
Austrian Federal Ministry of Education, Science and Research (FWF) and
FWF Austrian Science Fund No.~P~31361-N36;
National Key R\&D Program of China under Contract No.~2022YFA1601903,
National Natural Science Foundation of China and research grants
No.~11575017,
No.~11761141009, 
No.~11705209, 
No.~11975076, 
No.~12135005, 
No.~12150004, 
No.~12161141008, 
and
No.~12175041, 
and Shandong Provincial Natural Science Foundation Project ZR2022JQ02;
the Czech Science Foundation Grant No. 22-18469S;
Horizon 2020 ERC Advanced Grant No.~884719 and ERC Starting Grant No.~947006 ``InterLeptons'' (European Union);
the Carl Zeiss Foundation, the Deutsche Forschungsgemeinschaft, the
Excellence Cluster Universe, and the VolkswagenStiftung;
the Department of Atomic Energy (Project Identification No. RTI 4002), the Department of Science and Technology of India,
and the UPES (India) SEED finding programs Nos. UPES/R\&D-SEED-INFRA/17052023/01 and UPES/R\&D-SOE/20062022/06; 
the Istituto Nazionale di Fisica Nucleare of Italy; 
National Research Foundation (NRF) of Korea Grant
Nos.~2016R1\-D1A1B\-02012900, 2018R1\-A2B\-3003643,
2018R1\-A6A1A\-06024970, RS\-2022\-00197659,
2019R1\-I1A3A\-01058933, 2021R1\-A6A1A\-03043957,
2021R1\-F1A\-1060423, 2021R1\-F1A\-1064008, 2022R1\-A2C\-1003993;
Radiation Science Research Institute, Foreign Large-size Research Facility Application Supporting project, the Global Science Experimental Data Hub Center of the Korea Institute of Science and Technology Information and KREONET/GLORIAD;
the Polish Ministry of Science and Higher Education and 
the National Science Center;
the Ministry of Science and Higher Education of the Russian Federation
and the HSE University Basic Research Program, Moscow; 
University of Tabuk research grants
S-1440-0321, S-0256-1438, and S-0280-1439 (Saudi Arabia);
the Slovenian Research Agency Grant Nos. J1-9124 and P1-0135;
Ikerbasque, Basque Foundation for Science, and the State Agency for Research
of the Spanish Ministry of Science and Innovation through Grant No. PID2022-136510NB-C33 (Spain);
the Swiss National Science Foundation; 
the Ministry of Education and the National Science and Technology Council of Taiwan;
and the United States Department of Energy and the National Science Foundation.
These acknowledgements are not to be interpreted as an endorsement of any
statement made by any of our institutes, funding agencies, governments, or
their representatives.
We thank the KEKB group for the excellent operation of the
accelerator; the KEK cryogenics group for the efficient
operation of the solenoid; and the KEK computer group and the Pacific Northwest National
Laboratory (PNNL) Environmental Molecular Sciences Laboratory (EMSL)
computing group for strong computing support; and the National
Institute of Informatics, and Science Information NETwork 6 (SINET6) for
valuable network support.


\begin{thebibliography}{99}
\bibitem{chargeconjugate}
The inclusion of the charge-conjugate decay mode is implied throughout the letter unless otherwise stated.
\end{thebibliography}
\end{document}